# Seeking celestial Positronium with an OH-suppressed diffraction-limited spectrograph

GORDON ROBERTSON,[1,2,*] SIMON ELLIS,[2] QINGSHAN YU[1,3],
JOSS BLAND-HAWTHORN,[1,3] CHRISTOPHER BETTERS,[1,3] MARTIN ROTH,[4]
SERGIO LEON-SAVAL[1,3]

[1] *Sydney Institute for Astronomy, School of Physics, University of Sydney, NSW 2006, Australia*
[2] *Australian Astronomical Optics, Macquarie University, 105 Delhi Rd, North Ryde, NSW 2113, Australia*
[3] *Sydney Astrophotonics Instrumentation Laboratory, University of Sydney, NSW 2006, Australia*
[4] *innoFSPEC Potsdam, Leibniz Institute for Astrophysics Potsdam (AIP), An der Sternwarte 16, 14482 Potsdam, Germany*
*Corresponding author: Gordon.Robertson@sydney.edu.au

**Celestially, Positronium (Ps), has only been observed through gamma-ray emission produced by its annihilation. However, in its triplet state, a Ps atom has a mean lifetime long enough for electronic transitions to occur between quantum states. This produces a recombination spectrum observable in principle at near IR wavelengths, where angular resolution greatly exceeding that of the gamma-ray observations is possible. However, the background in the NIR is dominated by extremely bright atmospheric hydroxyl (OH) emission lines. In this paper we present the design of a diffraction-limited spectroscopic system using novel photonic components - a photonic lantern, OH Fiber Bragg Grating filters, and a photonic TIGER 2-dimensional pseudo-slit - to observe the Ps Balmer alpha line at 1.3122 μm for the first time.**

## 1. INTRODUCTION

It has long been known that positrons are produced in the Milky Way galaxy, through the detection of the 511 keV γ-rays produced as the positrons annihilate with ambient electrons [1,2,3]. The emission is concentrated towards the Galactic Center, but has wide extension in the bulge and (at lower intensity) along the plane of the Galaxy [3].

However, the precise nature and location of the source(s) of the positrons are unclear. The γ-ray distribution is consistent with a population of discrete sources in the Galactic bulge. Microquasars consist of an accretion disk surrounding a stellar-mass black hole and are a possible source because annihilation γ-rays have been detected from one of them [4]. Alternatively, the positrons could arise from objects such as exploding stars (supernovae) or processes occurring near the Galactic Center Black Hole itself (with subsequent diffusion through the interstellar medium before annihilation). More exotic sources could include annihilating dark matter [3]. The main reason for this uncertainty is the coarse angular resolution (~2°) of the best available γ-ray observations. It is this problem that we aim to overcome with the present project.

Before annihilation, the positrons and electrons briefly combine to form positronium (Ps) atoms, in one of two spin states. In the singlet state (para-positronium) the electron and positron have anti-parallel spins, and the system undergoes rapid decay into two oppositely- directed 511 keV γ-rays. But the triplet state (ortho-positronium), with parallel spins, has a much greater mean life of 142 ns in the ground state and 3.8 μs in the n=3 quantum state. It decays into three γ-rays, which results in a wide range of photon energies rather than the identifiable sharp 511 keV line [5]. However, the longer lifetime allows for the possibility of optical transitions from excited states before annihilation [2]. The energy levels for positronium are almost exactly half those of hydrogen, resulting in spectral lines with twice the wavelength of the hydrogen spectrum.

The best candidate for observation using ground-based telescopes is the analog of hydrogen Balmer α (656 nm), with Ps Balmer α (n = 3→2) having a wavelength of 1.3122 μm. However, this wavelength lies in the band subject to strong OH emission from the Earth's atmosphere. The emission is variable in time and across the sky, making sky subtraction after detection unsuccessful. Furthermore, the OH lines are so strong that any desired spectral features can be swamped by a nearby OH line in the wings of the instrumental response function (Line Spread Function, LSF). Figure 1 shows the atmospheric emission in the region of interest [6].

We propose to overcome this problem using Fiber Bragg Grating (FBG) technology to suppress the OH lines at high resolution, and then to observe the relevant spectral region with a small spectrograph. Our proposal is based on the Photonic Integrated Multi-Mode Spectrograph (PIMMS) concept [7,8], which enables a small inexpensive spectrograph to receive the light from any telescope, while at the same time gaining the advantages from photonic technology, such as FBGs.

Observing in the NIR does have a crucial advantage over the visible band, namely the greater transmission through dust obscuration, which is prominent towards the Galactic Center. It also benefits from the availability of adaptive optics systems at major telescopes. In any case, there are no Ps lines in the visible range, with its Lyman series lying in the UV.

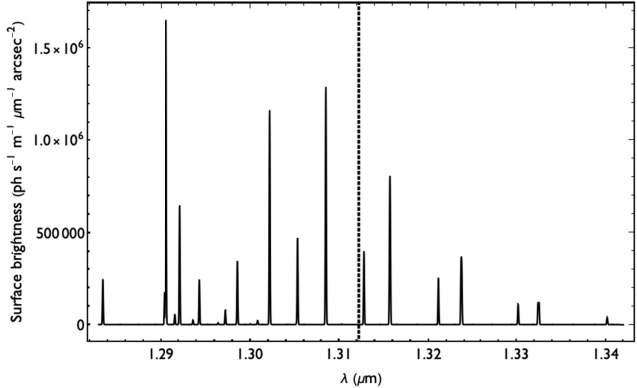

Fig. 1. Atmospheric OH emission spectrum [6]. The dashed vertical line shows the wavelength of positronium Balmer α emission for a source with zero radial velocity relative to the Earth. The OH lines are shown with width Δλ equal to the 0.1 nm expected width of the FBG notches (*i.e.* λ/Δλ = 13,100).

## 2. INSTRUMENT CONCEPT

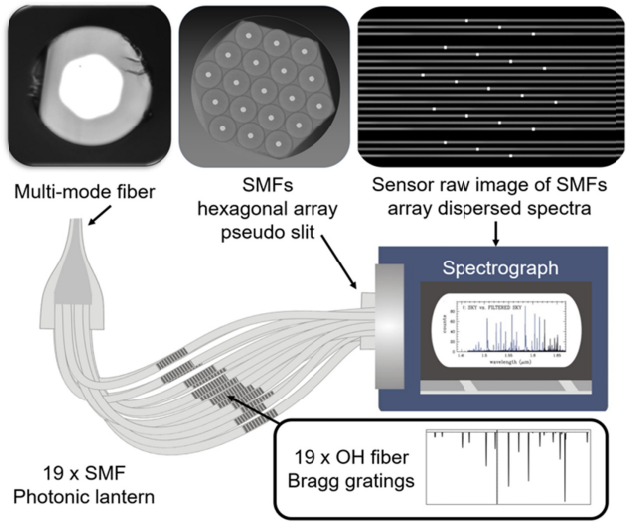

Fig. 2 . Schematic of the proposed system. The target is imaged by a telescope to a multi-mode fiber, which transitions in a photonic lantern to 19 single-mode fibers, each of which has a custom Fiber Bragg Grating to reject specific wavelengths of OH lines. The SMFs are arranged in a hexagonal pattern at the entrance plane of a small diffraction-limited spectrograph, with spectra imaged on a C-RED-One detector.

Figure 2 shows a schematic of the proposed system. Objects will be imaged by a telescope on to the multi-mode fiber (MMF) end of a photonic lantern, which transitions the light into 19 single-mode fibers (SMF) [9,10,11]. In each SMF a custom FBG will be imprinted, with sharp and deep rejection notches at the wavelengths of the bright OH lines near 1.312 µm. This technology has been demonstrated to successfully suppress the OH emission [12,13,14].

The advantage of the FBG approach is that the atmospheric OH lines are very narrow and are also stable in wavelength – so rejection notches at high resolving power can remove them, while losing a small fraction of the flux of the desired object, which is observed at lower resolving power. The FBGs are temperature-stabilised to ensure that the notch wavelengths remain aligned with the atmospheric emission lines. Importantly, the notch profile can be designed with steep sides and a flat bottom – which for this purpose is superior to the typical LSF of a high-resolution spectrograph. For example Trinh et al. [15] demonstrated notches of 28 dB depth, resolving power 10,400 and square sides. It will only be necessary to suppress some 6-10 OH lines for the present project, which will enable the FBGs to be configured to give narrower notches than would be the case for FBGs seeking to suppress OH over a wider band.

In an OH suppression spectrograph the resolving power is generally determined by the expected line widths for the target, and not by the requirement to resolve or separate the OH lines themselves, since these have been suppressed at higher resolution by the FBGs. For Ps forming through radiative combination of positrons with free thermalized electrons, the line width Δλ is given by [16]:

$$\frac{\lambda}{\Delta\lambda} = 73774.4\, T^{-0.44}$$

so for typical interstellar medium temperatures $T = 10^3 – 10^5$ K, the resolving power λ/Δλ required is $R \approx 500 – 3500$. However, in the present case the Ps line is extremely faint compared to the OH lines, and therefore we wish to ensure the Ps signal is well separated from the residual OH lines. The nearest OH line is at 1.31278 µm, separated by 0.58 nm from Ps Balmer α. A resolving power of $R \approx 4500$ ensures a separation of two resolution elements, and will adequately resolve any Ps line.

Since the light emerges in single-mode fibers, we can obtain good resolving power with a small diffraction-limited custom spectrograph. A diffraction-limited spectrograph with an SMF input offers two important advantages for this experiment. First, it allows a very stable and compact spectrograph at moderate resolving power (R~4000) to be built for a relatively low cost [17]. Secondly, it provides a spectrograph line spread function with very low scattering [18], which allows clean sky-subtraction – still an important consideration for faint signals even with OH suppression [13]. This latter advantage outweighs the increase in detector noise due to the multiple SMF inputs spreading the light over more pixels than the equivalent MMF injection (as used in previous OH suppression instruments), especially given the low read-out noise of the detector.

The detector will be a C-RED-One Avalanche Photodiode (APD) array[1] [19]. This has the crucial advantage of sub-electron readout noise, meaning it is effectively a photon-counting array. It also has high detection quantum efficiency and low dark noise. The array consists of $320 \times 256$ pixels, at a pitch of 24 µm.

The limited number of pixels does influence the design of our instrument: if the fibers were arranged along a linear pseudo-slit, at the minimum feasible separation of ~60 µm (due to engineering limitations in the fabrication; assuming etched 80µm fiber such as the Thorlabs SM1250G80), the slit length becomes 1.14 mm, and fitting the full slit length on the detector places an upper limit on the magnification provided by the collimator/camera combination. The result is that in the spectral direction the profiles would be severely undersampled, at ~1.35 pixels/Full Width at Half Maximum (FWHM). We avoid this by arranging the fibers in a hexagonal pattern at the spectrograph 'slit' plane, in what has become known as the Photonic TIGER pattern [20,21]. The result is an effective slit length of 0.3 mm, allowing a higher magnification and optimal sampling of the LSF. This technology has been demonstrated in a compact high-resolution diffraction-limited Raman spectrograph [22] using multicore fiber, but the principle is also valid for a bundle of individual fibers as proposed here. Furthermore, the sensor size limits this approach to a small number of SMF inputs, hence we use a 19 SMF photonic lantern, although with larger sensors the number of SMFs could be increased. The 2-dimensional arrangement instead of a linear pseudo-slit results in different wavelength ranges on the detector for different fibers, and a consequent reduction in the wavelength range that is common to all fibers. The common range is still sufficient for our purpose, however. Having the individual spectra staggered in this way limits the use of on-chip binning along the spatial axis, but binning is in any case unnecessary with a nearly noise-free detector.

### A. Spectrograph design

In the first instance, we have sought a spectrograph design using only commercial off-the-shelf components. Table 1 lists the selected components. All lenses are anti-reflection coated for 1.05 – 1.7 µm. The transmission diffraction grating is of Volume Phase Holographic (VPH) type. Figure 3 shows the optical layout. The main design problem is the collimator, which is quite fast: $f/4.5$ to the $1/e^2$ points, but about $f/3.4$ to avoid excessive truncation of the Numerical Aperture (NA) = 0.11 Gaussian beams [23]. The 3-lens combination for the collimator has an effective focal length of 61 mm, giving a nominal beam diameter of 13.5 mm to the $1/e^2$ points.

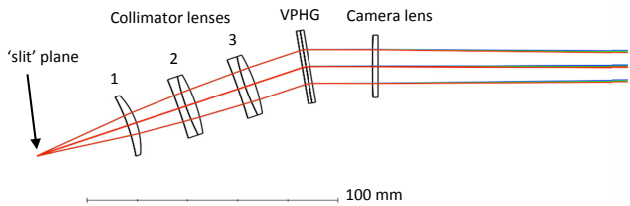

Fig. 3. Front-end of the spectrograph. The injection from the fibers is at the left-hand side. The beam encounters the 3-element collimator, VPH grating, and the camera lens. The detector is off the diagram to the right. The rays shown are for NA = 0.11, from the on-axis fiber.

Figure 4 shows the detector plane as well as the window and 4 filters (including cold-stop) of the C-RED-One.

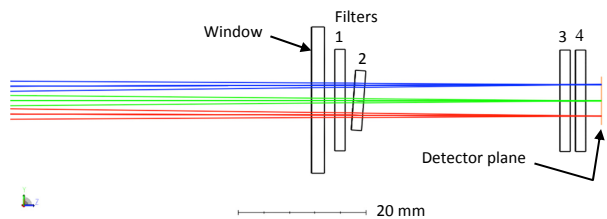

Fig. 4. The back-end of the spectrograph, showing the warm sapphire Dewar window and 4 cooled Hereaeus Infrasil 302 filters of the C-RED-One detector. The detector plane is at the right-hand side. The same rays as in Fig. 3 are shown, with wavelengths 1.299 µm (blue), 1.312 µm (green) and 1.325 µm (red). The cold stop is at filter 2 and has a diameter of 9.5 mm.

**Table 1. Components of the spectrograph**

| Item | f /mm | Diam /mm | Type | Part description |
|---|---|---|---|---|
| Slit | | | Hex pattern | 19 single-mode fibers |
| Collim lens 1 | 100 | 25.4 | meniscus | Thorlabs LE1234-C |
| Collim lens 2 | 200 | 25.4 | achr doublet | Thorlabs AC254-200-C |
| Collim lens 3 | 200 | 25.4 | achr doublet | Thorlabs AC254-200-C |
| Grating | | 30 | VPH | Wasatch 250 l/mm at 1250 nm |
| Camera lens | 748 | 25.4 | plano-convex | Thorlabs LA1978-C |
| Detector | | | | C-RED One, 320 × 256, 24µm |

## 3. SPECTROGRAPH PERFORMANCE

### A. Format

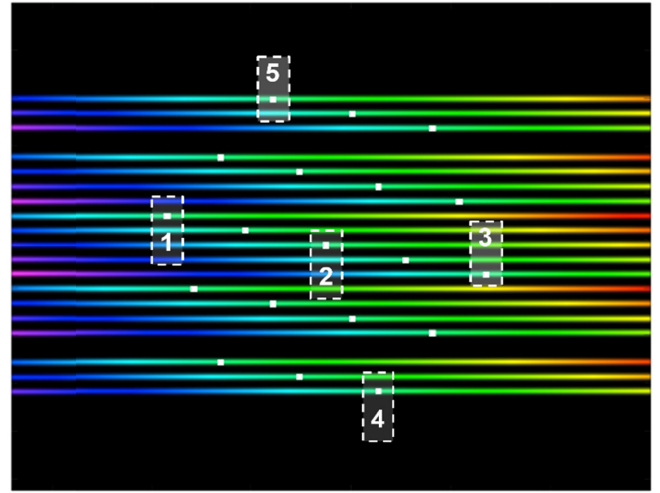

Fig. 5. Pseudo-color representation of the spectra, showing the full detector area. The white squares show the imaged location of each fiber at the center wavelength of 1.312 µm, in the inclined hexagonal Photonic TIGER pattern; numbered squares refer to the rows in Figure 6 and Table 2. The individual spectra are shown with colors ranging from magenta (1.283 µm) to red (1.342 µm), and with the calculated Gaussian profile in the spatial direction. Detector pixelization is not shown here. The Figure shows that the layout does not use the full extent of the detector's spatial axis.

Figure 5 shows a simulation of the format of the spectra on the detector. The hexagonal pattern of the fibers at the injection plane has a pitch of 80 µm, and is inclined by 10.9° so that spectra do not overlap [21]. The pattern is magnified by the ratio of the camera and collimator focal lengths, which is 12.3.

### B. Image quality

Figure 6 shows ray-trace spot diagrams at the detector plane, for the center and four extremes of the hexagonal pattern, at the center wavelength plus the extremes of the common range. The actual beams are Gaussian, and will be shown below. But for assessing aberration performance, it is useful to first examine the ray-trace results using a beam with uniform apodisation, and NA = 0.11. (The actual Gaussian beams will be tapered to 13.5% of peak intensity at the same angle as the edge of the uniform beam, but will then extend somewhat beyond.)

Figure 6 shows all spots are well within the Airy disk, *i.e.* all images are close to diffraction-limited. They show good correction of spherical aberration, but do have a little longitudinal chromatic aberration, as expected from the singlet meniscus lens and camera lens.

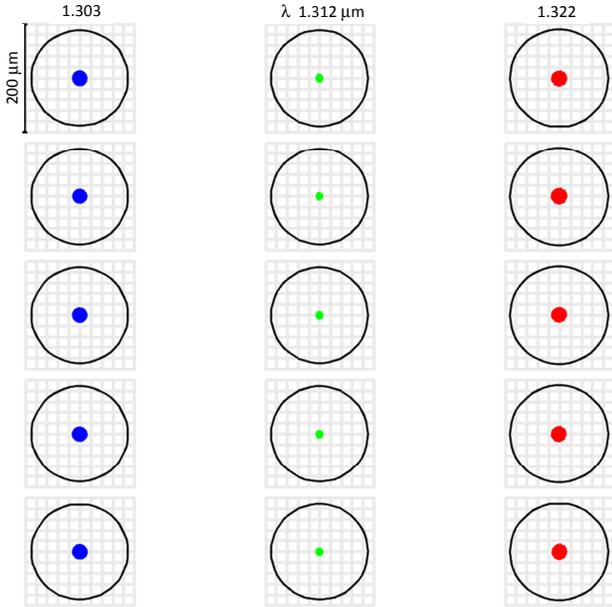

Fig. 6. Matrix spot diagram showing ray trace results at the center wavelength (middle column) and the two extreme wavelengths of the common range. The five rows show the central fiber and the four with the greatest displacements in both the spatial and wavelength directions. The slit plane XY positions of the inputs for each row are shown in Table 2 (and annotated in Figure 5 at the center wavelength). The circles show the Airy disk radius; the box side is 200 µm. These plots assume rays are uniform within NA = 0.11. The individual traced rays have merged into a single circular patch in each case, because the aberrations are minimal.

Table 2. Slit plane XY positions of spot diagram fibers

| Row no. | $X_{slit}$ /mm | $Y_{slit}$ /mm |
|---|---|---|
| 1 | 0.0303 | -0.1571 |
| 2 | 0 | 0 |
| 3 | -0.0303 | 0.1571 |
| 4 | -0.1512 | 0.0524 |
| 5 | 0.1512 | -0.0524 |

ZEMAX Physical Optics Propagation[2] has been used to find the image properties when using Gaussian beams with NA = 0.11 from the fiber cores. The 2D images in the focal plane were found, then exported to MATLAB[3] where they were projected to the wavelength axis, giving the simulated LSF. MATLAB was also used to find the accurate FWHM values for the LSFs.

Figure 7 shows the projected LSFs for two cases – the on-axis fiber at the center wavelength, and fiber no. 3 in Figure 5 and Table 2, observed near the detector edge at 1.325 µm. . As expected from the spot diagrams in Figure 6, there is very slight broadening for the off-axis case. Both have a clean Gaussian profile.

The FWHMs are 57.9 µm on axis and 60.2 µm for the other. Thus there is a 4% broadening away from the detector center. The resolving power ($\lambda$/FWHM) is 4410 for the on-axis case, and 4240 for the off-axis case. With 24 µm pixels, the detector sampling is 2.4 – 2.5 pixels/FWHM, a near-ideal balance between accurate characterization of spectral features and wavelength coverage.

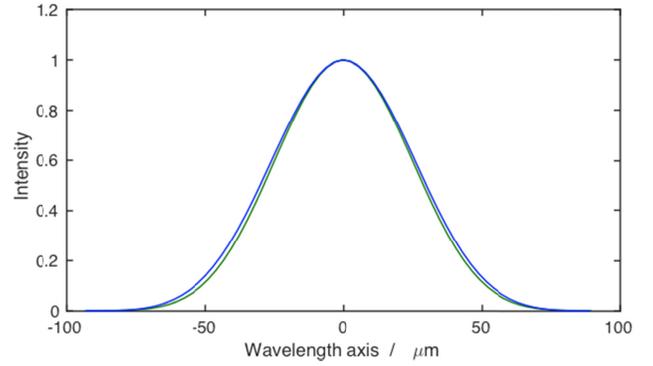

Fig. 7. Monochromatic profiles of the Line Spread Function as calculated using Gaussian beam inputs, for the center of the detector (green curve, location '2' in Fig. 5) and off-center (blue curve, fiber '3', at $\lambda$ 1.325µm).

### C. Wavelength range

Figure 8 shows the number of fiber spectra available at any given wavelength. Coverage begins at 1.2827 µm, with one fiber, and ends at 1.3420 µm, again with one fiber. Within the common range 1.303 – 1.322 µm, all 19 spectra are on the detector. The common range is equivalent to a radial velocity range of -2070 to +2270 km/s for a spectral line with rest wavelength 1.312 µm. This will be adequate for any source within the Milky Way galaxy, but for extragalactic sources such as Active Galactic Nuclei it extends only to redshift 0.0076. (However, coverage with fewer spectra extends to 6700 km/s or redshift = 0.023.) If desired, the Ps line wavelength of 1.312 µm could be moved towards the shorter wavelength end of the detector by changing the grating and camera tilt angles, so increasing the available range of redshifts, at the expense of coverage shortward of the line.

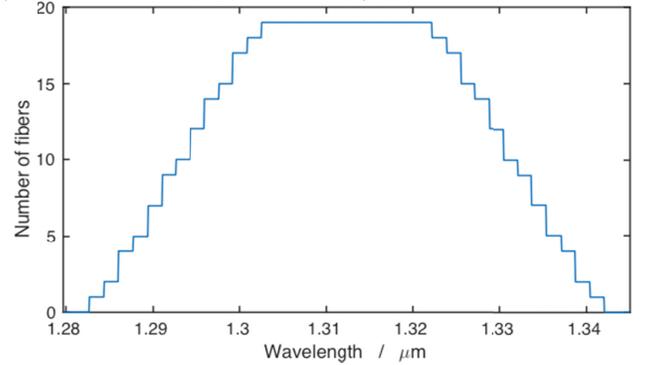

Fig. 8. The number of individual fiber spectra available at each wavelength. Only 1.303 – 1.322 µm is common to all 19 fibers.

## 4. INSTRUMENT PERFORMANCE

### A. On-sky angular capture

Due to atmospheric turbulence ("seeing") a ground-based telescope is not in general diffraction-limited, and hence cannot efficiently couple a star image into a single-mode fiber. For this reason our system uses a multi-mode fiber which transitions to 19 SMFs. We now evaluate the angular size of the telescope's seeing disk that can be captured by our 19-mode system.

For a Gaussian beam the far-field beam divergence half-angle (to the $1/e^2$ point) is

$$\theta = \frac{\lambda}{\pi r_0}$$

[e.g. 24], where $r_0$ is the mode-field radius. For moderate NAs such as our value of 0.11, θ (in radians) is approximately equal to the NA. For the individual SMFs we therefore have

$$r_0 NA_{fib} = \frac{\lambda}{\pi}$$

We will make use of the principle of etendue (AΩ, area × solid angle) conservation. The AΩ product will be increased over that for a single fiber by a factor equal to the number $N$ of fibers (modes) included. Since the above equations relate to radii and plane angles rather than areas and solid angles, the increase will be a factor of $\sqrt{N}$.

Consider a telescope with primary mirror diameter $2R$, forming star images with angular FWHM Γ arcsec (the usual measure of astronomical seeing). Seeing disks have approximately Gaussian profiles, so a meaningful comparison of AΩ products can be made.

For a Gaussian the $1/e^2$ points are separated by $1.699 \times$ FWHM [23]. Expressing the seeing angle as an effective NA of the telescope,

$$NA_{tel} = \frac{1.699\,\Gamma}{2} \times \frac{1}{3600} \times \frac{\pi}{180}$$

Etendue conservation is expressed as

$$\sqrt{N}\, r_0\, NA_{fib} = R\, NA_{tel}$$

which leads to

$$\Gamma = 7.73 \times 10^4\, \frac{\sqrt{N}\,\lambda}{R}$$

where Γ is in arcseconds.

This is the seeing disk diameter that could be efficiently coupled into our system. For R = 2 (i.e. a 4-metre telescope) and $N$ = 19, we find Γ = 0.22 arcseconds. This is smaller (better) than the natural seeing at most observatories. For an 8-metre telescope good coupling requires seeing of 0.11 arcsec. In both cases an adaptive optics system would in general be needed to achieve good coupling, but without it we can still observe using the core region of a natural seeing disk. A larger detector would enable us to use a larger photonic lantern, e.g. 37 SMFs, which would increase the angular capture (field of view) of our instrument.

### B. Sensitivity estimate

We have made preliminary estimates of the signal/noise (S/N) ratio obtainable from microquasar point sources of Ps Balmer α radiation, using estimated positron fluxes from [2].

The instrument throughput depends strongly on the coupling efficiency achieved at the injection to the MMF. This in turn depends on the seeing and can be greatly improved through the use of adaptive optics. At the present early design stage, we use indicative figures to estimate the overall system efficiency. Experience with the GNOSIS system [15] shows that throughput of the lantern/FBG is ~70%; lenses are all anti-reflection coated, with ~0.5% loss, the VPH grating has efficiency ~90%, and the detector quantum efficiency is at least 70%. Combining with telescope throughput we have an indicative system efficiency of 0.38 (excluding coupling losses).

The OH suppression improves the S/N by a factor of 1.5 - 2 for typical sources, although this is sensitive to the relative brightness of the Ps line versus the source continuum. The reason for the improvement is the reduction in the night sky background - the fainter the source the greater the relative improvement. If a target source had a very bright continuum, the OH suppression would be no advantage, due to the extra losses from the FBG unit, although the diffraction-limited LSF would still provide a major benefit by making sky subtraction more accurate. However, our likely target sources are faint, and will benefit from OH suppression.

Note that we seek to maximise S/N - not simply flux from the source. The optimum field of view therefore depends on the site conditions (i.e. seeing), and the level of adaptive optics correction, as well as the throughput of the photonic lanterns, which is a function of injected focal ratio [25]. The best S/N is achieved by injecting the maximum target light into the smallest possible field of view, so minimising the night-sky contamination of the signal.

With a 4 m telescope, assuming seeing of 0.5 arcsec and instrument + telescope throughput of 0.38, we obtain S/N in a 10-hour observation equal to 7 for the brightest example (XTE J1118+480). Figure 9 shows a simulated observation for a similar source and demonstrates the value of the OH suppression. For other sources towards the Galactic Center region, S/N of 5-6 could be expected in 10 hr. The OH suppression is crucial for such objects. As noted, our system would be ideally suited for use behind an adaptive optics facility at an 8 m telescope, which would greatly increase the coupling by matching the delivered image size to our system's angular capture, without increasing the amount of night-sky light. Fainter objects would then be detectable.

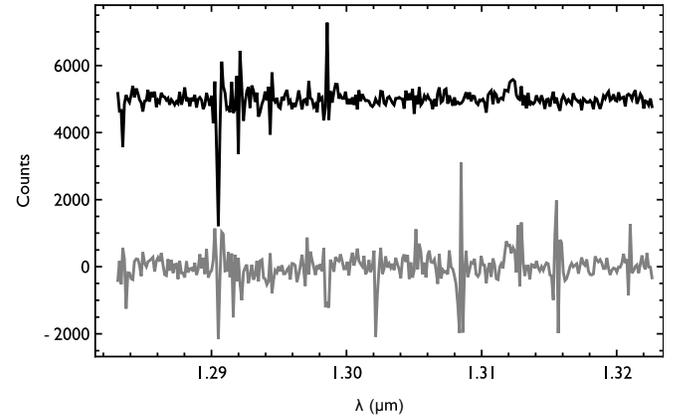

Fig. 9. Simulation of a source with an incident Ps Balmer α flux of 50 $ph/s/m^2/\mu m$, with continuum of J=18 mag, in 10 hrs on a 4 m telescope. The OH suppression notches are assumed to have depth 30dB and width 100 pm. The result has S/N = 8 when OH suppressed (black), which is 2.1× better than the non-suppressed case (gray). The positronium spectral line is the peak at 1.312 μm, and the vertical gray lines mark the position of suppressed OH lines. The imperfect subtraction of these lines in the non-suppressed case is clearly detrimental to the detection.

## 5. CONCLUSIONS

We have presented a feasible concept for a small instrument to be attached to a telescope to search for NIR recombination line emission from positronium. This paper outlines the photonic technologies employed, concentrating on the optical design of the proposed spectrograph.

The limited angular capture means that our instrument is suited to detection of recombination radiation from point sources but not truly diffuse regions. However, an upper limit from a non-detection can still provide discrimination between point sources and diffuse emission, which is an important test, both for truly diffuse emitters of positrons (such as dark matter) and of positron transport - do positrons annihilate close to their sources or not?

We are working towards completion of the design, to be followed by construction and on-sky use initially at the Anglo-Australian Telescope,

which is situated at an ideal latitude for observations of the Galactic Center region.

**Funding sources and acknowledgments**. MR acknowledges financial support from the German federal ministry for education and science (BMBF) grant 03Z22AI1. QY acknowledges scholarship support from the Faculty of Science Dean's International Postgraduate Research Scholarship. We thank the reviewers for comments which have improved the paper.

**Disclosures.** The authors declare no conflicts of interest.

---

[1] www.first-light-imaging.com

[2] www.zemax.com

[3] www.mathworks.com